\documentstyle[aps,graphicx]{revtex}
\newcommand{\bm}[1]{\mbox{\boldmath $#1$}}
\newcommand{\sv}{\mbox{\boldmath $\hat{\sigma}$}}

\begin{document}

\title{Andreev reflection in unitary and non-unitary triplet states}
\author{Carsten Honerkamp and Manfred Sigrist
\footnote{New address: Yukawa Institute for Theoretical Physics, Kyoto University, Kyoto 606-01, Japan.}}
\address{Institute of Theoretical Physics, ETH H\"onggerberg, CH-8093 Z\"urich,
Switzerland}
\maketitle

\begin{abstract}
The quasiparticle reflection and transmission properties at normal
conductor-superconductor interfaces are discussed for unitary and
non-unitary spin triplet pairing states recently proposed for Sr$_2$RuO$_4$.
We find resonance peaks in the Andreev reflection amplitude, which are related
to surface bound states in the superconductor. They lead to 
conductance peak features below the quasiparticle gap in the
superconductor. The position of the peaks depends on the angle of
the incident quasiparticles in a different way for unitary and
non-unitary states. Based on this observation we propose a possible
experiment which allows to distinguish between the two kinds of
superconducting states.   
\end{abstract}
 
\maketitle

\vskip2pc

\section{Introduction}
The recent observation of superconductivity in Sr$_2$RuO$_4$ has attracted
much attention because of the close structural
similarity with some of the high-temperature superconductors\cite{mae}. 
The critical temperature, $ T_c \sim 1 $K , is, however, 
rather low. In fact there is little similarity between 
 Sr$_2$RuO$_4$ and high-temperature superconductors beyond the crystal
structure. In strong contrast to the CuO$_2$-systems  Sr$_2$RuO$_4$ is
a good metal as a stoichiometric compound, displaying clear Fermi
liquid properties. All three $ t_{2g} $-d-orbitals of Ru$^{4+}$
($4d^4$) yield bands which cross the Fermi energy and lead to two
electron-like and one hole-like Fermi surfaces observed 
in the beautiful de Haas-van Alphen experiments of Mackenzie et al.\cite{mack}. The layered
structure 
of the system leads to a comparatively small dispersion of the bands
along the c-axis that is reflected in the highly anisotropic
resistivity. Therefore  we find an essentially  
two-dimensional electron system and the nearly cylindrical Fermi surfaces are open along
the c-axis. There are various
indications for strong correlation effects. Among others the indication of  a
considerable mass enhancement, $ m^* \sim 4 m_0 $, has been
found\cite{mae}, which 
causes an enhanced linear specific heat coefficient $\gamma$. Also
the Pauli spin susceptibility is enhanced. 

Sr$_2$RuO$_4$ is a member of a homologous series of compounds,
Sr$_{n+1}$Ru$_n$O$_{3n+1}$, where $ n $ is the number of RuO$_2$-layers
per unit cell. The single layer compound Sr$_2$RuO$_4$ seems to have
unique properties among the members investigated so far. 
Both the compound with $ n=2 $ (double layer) and $ n = \infty $
(infinite layer) are itinerant ferromagnets. 
This suggests that also
in the single layer compound ferromagnetic spin fluctuations should be
very important, again in contrast to the CuO$_2$-compounds where
antiferromagnetic spin fluctuations are dominant.

The presence of strong correlation effects implies that it is 
difficult for electrons to form Cooper pairs in the usual s-wave
channel. Thus a higher relative angular momentum would be more favorable. 
There are also interesting
similarities with $ ^3$He which led to the suggestion that Cooper
pairing in the spin triplet (odd-parity)  configuration might be
favored\cite{siri}. Clearly also the presence of strong ferromagnetic spin
fluctuations and the Hund's 
rule coupling of the d-orbitals on the Ru-anion point into the same
direction. 

At present, there is no unambiguous experimental indication for triplet
pairing. Nevertheless, there is strong evidence
for unconventional superconductivity. The superconducting state is
extremely sensitive to disorder. Only very clean samples show
superconductivity and a small concentration of non-magnetic
impurities (Al) suppresses
the transition temperature to zero. The nuclear quadrupolar relaxation
(NQR) deviates from that of conventional
superconductors\cite{nqr}. 
It does not 
have any sign of a Hebel-Slichter peak and shows a low temperature
Korringa-like behavior. 
Furthermore, the specific heat data indicate a large residual density
of states in the superconducting phase whith a linear-$T$ coefficient 
$ \gamma $ with about half of the normal state value even in the
purest samples\cite{lincv} consistent with NQR.

The number of possible spin triplet Cooper pairing states is
large and the question arises whether any of those states could explain 
the presence of the large residual density of states.
One proposal was a non-unitary state which can only occur for spin
triplet pairing\cite{siz}. For such a state the
quasiparticle excitation spectrum consists of two branches, one of
which could be gapless leading to 
finite density of states at zero energy which is half the normal state
value. The nature of this state is very similar to the $ A_1 $-phase of
superfluid $ ^3$He which is stabilized in a magnetic field. It is
still an open problem whether there is a mechanism that could give rise to
such a state in zero external magnetic field. 
An alternative explanation which avoids this difficulty 
was recently proposed by Agterberg et
al.\cite{agt} based on the fact that there are three distinct Fermi surface
sheets. For symmetry reasons two sheets are nearly decoupled from the
third one for Cooper pair scattering. Therefore the
superconducting gap would have different values on different sheets. This
leaves space for rather low-lying quasiparticle excitations on certain Fermi
surfaces which would appear as residual density of states in the
specific heat measurement.
There is no clear experimental distinction between the two proposals
so far.

There is good 
experimental evidence that also the heavy fermion superconductor UPt$_3$
has spin-triplet pairing ($ T_c \approx 0.5$K) \cite{kitaoka}. In this case
  two distinct superconducting phases, a low- and a high-temperature
  phase, occur. The low-temperature phase has presumably broken time
  reversal symmetry and may also have a non-unitary state \cite{luke,sauls}.
The discussion we do here for Sr$_2$RuO$_4$ may also apply to this
compound to some extent.

In this paper we investigate the quasiparticle tunneling from a
normal metal to the spin triplet superconductor (NS) for both the
unitary and non-unitary pairing state. This type of problem was first
analyzed by Bruder many years ago \cite{bruder}. We would like to
focus here on the interesting aspect of broken time reversal symmetry in
the non-unitary state. This property influences the
presence of so-called Andreev bound states at the NS interface. We
will show that the energy of the bound states depends on the angle of
the incident quasiparticle (hole) which is a direct consequence of the
polarized internal angular momentum of the non-unitary state. These
bound states appear as anomalies in the current-voltage (IV) characteristics
of the interface. In general, measurements of the IV-characteristics
include an average over a wide range of  incidence angles. We will
show, however, that it is possible to introduce a certain selection of
angles by the method of magnetic focusing previously used for the
study of the Andreev effect in conventional superconductors \cite{ben}.

\section{The spin-triplet pairing states}

For spin-triplet superconducting states the Cooper pairs possess a
spin-1 degree of freedom. This state is described by a gap function which
is, in general, a $ 2 \times 2 $-matrix in spin space and can be 
parameterized by a vector function $ {\bf d}({\bf k}) $ 
\begin{equation}
 \hat{\Delta} ({\bf k}) =  i \sigma_2 ( {\bf d} ({\bf k}) \cdot \sv
)  
\end{equation}  
which is
odd in $ {\bf k} $  \cite{leg,su}.
The energy spectrum of the superconducting quasiparticles is
 given by 

\begin{equation} 
E^{\pm }_{\bf k} = \sqrt{ \epsilon^2 ({\bf k }) + | {\bf d} ({\bf 
    k}) |^2 \pm |{\bf q} ({\bf k})| } \label{spectrum}
\end{equation} 
where the vector ${\bf q}({\bf k})= i {\bf d} \times {\bf d}^*$. $\epsilon( {\bf k})$ is the band energy measured with respect to the chemical potential. It is
$ {\bf q} $ which distinguishes between two classes of spin-triplet
pairing states, the unitary and non-unitary state. If $ {\bf q}=0 $
the state is 
unitary and the two branches $ E^{\pm}_{\bf k} $ are identical. On the
other hand, for $ {\bf q} \neq 0 $ the state is called non-unitary. It
has two distinct quasiparticle energy spectra and breaks time-reversal
symmetry.

For simplicity we will restrict ourselves to the case of
a single cylindrical symmetrical Fermi surface and analyze an example
of both types of spin-triplet pairing states. These states result from
a classification of pairing states
including spin-orbit coupling and crystal field effects
\cite{siri,siz}.

The example for the non-unitary state has the form

\begin{equation} {\bf d} ({\bf k}) \propto  ( \hat{{\bf x}} + i
  \hat{{\bf y}} ) \, ( k_x - i k_y) \, , \label{sz} \end{equation} 
or
\begin{equation} 
\hat{\Delta } ({\bf k})  =   \Delta_0 \,  \left( 
\begin{array}{cc}k_x - i k_y & 0 \\ 0 & 0 \end{array} \right)  \, . 
\end{equation}
In this state, the Cooper pairs have $S_z= +1$ and $L_z= -1$
\cite{siz}. It is degenerate to a similar state
with opposite orientation of spin and angular momenta. The essential
feature of the two states is that the energy gap for one spin
projection of quasiparticles (spin-up for (\ref{sz}), $E^+_{\bf
  k}$-branch) has a constant magnitude, while it vanishes for the
branch $E^-$ on the whole Fermi surface as
can be seen immediately by inserting (\ref{sz}) into (\ref{spectrum}). As mentioned above
these gapless quasiparticles could account for the large residual
density of states. 

We consider the analogue of the Balian-Werthamer (BW) state\cite{bw} as an example for a unitary
state,
\begin{equation}  \hat{\Delta } ({\bf k}) = \Delta_0 \,  \left(
    \begin{array}{cc}k_x - i k_y & 0 \\ 0 & k_x + i k_y \end{array}
  \right)  \, . \label{bw2} \end{equation} 
Note that in two dimensions this is an equal spin pairing state and
has a gap with constant modulus for both spins on the cylindrical
Fermi surface. This state is stable in the weak coupling approach and 
could be the basis for the scenario of
orbital dependent superconductivity \cite{agt}.

\section{The model for the NS-interface}
\begin{figure}[t]\begin{center}
    \includegraphics[width=0.4\textwidth]{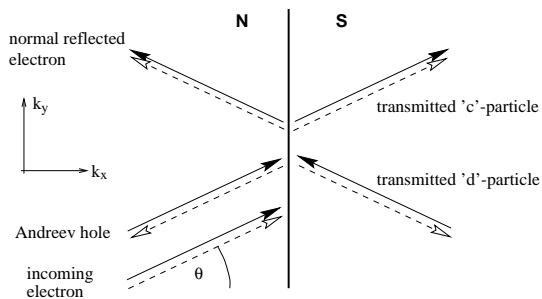} \end{center}  
\caption{Momenta (solid lines) and velocities (dashed lines) of the quasiparticles 
involved in the Andreev reflection amplitude $a(E, \theta)$.}
\label{gapvec} \end{figure}

In this section we analyze the properties of the
normal-metal-superconductor (NS) interface for the two types of pairing
states (\ref{sz}) and (\ref{bw2}). We apply the formulation by Blonder, Tinkham and Klapwijk
(BTK)\cite{btk} for NS interfaces. This allows us to derive
expressions for the reflection  probabilities of electrons.
The geometry of our problem contains the following restrictions.
The particles move in the $xy$-plane and the boundary between normal
metal ($x<0$) and superconductor ($x>0$) is the $yz$-plane located at $x=0$.  
We will use a step function for the spatial dependence of
the gap magnitude, $\hat{\Delta} ( {\bf k},{\bf x} ) = \Theta ( x )
\hat{\Delta} (\hat{{\bf k}}) $. The surface potential will be taken as
a delta potential 
 $U ( {\bf x})= H \delta (x)$. Then following BTK \cite{btk} the
 amplitudes for quasiparticle (Andreev-) reflection and transmission
 can be calculated using the appropriate boundary conditions for
the wave functions $\psi_{\mathrm{N}}$ on the
 normal and $\psi_{\mathrm{S}}$ on the superconducting side of the NS interface. 
 Since for triplet superconductors we have to take care of the spin
 structure of the gap function,  the quasiparticle wave functions are
 four-spinors in Nambu (particle-hole $\otimes$ spin) space. Their
 particle and hole components are 
 determined by the solutions of the Bogoliubov-deGennes
 equations \cite{su}, i.e. the diagonalization of the corresponding
 BCS mean field Hamiltonian \begin{eqnarray}   \hat{u}_{\bf k}  (
   \hat{E}_{{\bf k}} - \epsilon ({\bf k})  ) &=&   \hat{\Delta} ({\bf
     k})  
\,   \hat{v}_{-{\bf k}}^* 
 \, , 
\label{BdG} \\  
 \hat{v}_{-{\bf k}}^*  ( \hat{E}_{{\bf k}} + \epsilon ({\bf k})  ) &=&
 \hat{\Delta}^{\dagger} ( {\bf k}) \,   
\hat{u}_{\bf k}            
\nonumber     \, . \end{eqnarray}
Here $\hat{u}_{\bf k}$ and $\hat{v}^*_{-{\bf k}}$ are matrices in
spin-$\frac{1}{2}$ space, the Bogoliubov quasiparticle operators in the
superconductor are then given by  
\begin{eqnarray*} \bm{\gamma}_{{\bf k} \uparrow}^{\dagger} &=&
 v_{-{\bf k},11}^* {\sf c}_{-{\bf k} \uparrow } +    v_{-{\bf k},21}^*
 {\sf c}_{-{\bf k} \downarrow }      + 
 u_{{\bf k},11} {\sf c}_{{\bf k} \uparrow }^{\dagger} +    u_{{\bf
     k},21} {\sf c}_{{\bf k} \downarrow }^{\dagger}     \, ,   \\ 
 \bm{\gamma}_{{\bf k} \downarrow}^{\dagger}& =&
 v_{-{\bf k},12}^* {\sf c}_{-{\bf k} \uparrow } +    v_{-{\bf k},22}^*
 {\sf c}_{-{\bf k} \downarrow }      + 
 u_{{\bf k},12} {\sf c}_{{\bf k} \uparrow }^{\dagger} +    u_{{\bf
     k},22} {\sf c}_{{\bf k} \downarrow }^{\dagger}     \, .   
\end{eqnarray*}
For equal-spin pairing (ESP) states ((\ref{sz}) or (\ref{bw2})), the
spin-up and spin-down 
components decouple, therefore we obtain two-spinors with one
electron and one hole component. For triplet states, the Andreev hole
is a missing electron of the same spin projection as the incoming
particle. 
Following BTK \cite{btk}, we denote the
amplitude for Andreev reflection by $a$ and for normal reflection by
$b$. Moreover we introduce the amplitudes $c$ for transmission into
the superconductor as a Bogoliubov quasiparticle on the same branch of
the spectrum and $d$ for transmission with backscattering through the
Fermi surface, i.e. reverting the momentum $x$ component. The momenta
involved into the reflection processes are
shown in Fig.\ref{gapvec}.  
Now assume that the incoming particles are electron-like
quasiparticles with excitation energy $E$ with respect to the Fermi
energy, let their spins be polarized along the ''gapped'' direction
(in our convention spin-up). Then the wave function on the normal conducting side is the sum of the wave functions of the incoming particle and the reflected particles multiplied with their amplitudes,
\[ \psi_{\mathrm{N}} = \psi_{\mathrm{inc.}} +a \, \psi_{\mathrm{a}} + b \, \psi_{\mathrm{b}}\, ,  \] and on the superconducting side we have
\[ \psi_{\mathrm{S}} = c \, \psi_{\mathrm{c}} + d \, \psi_{\mathrm{d}} \, . \]
Limiting ourselves to sufficiently
small angles of incidence $\theta$ in the $xy$-plane, we can approximate all magnitudes of $x$
components of the involved momenta  by $k_x=k_F \cos \theta$. If we assume conservation of the quasiparticle momenta parallel to the boundary the
transverse wave function of all excitations on both sides is $e^{i k_F
  \sin \theta y }$. Writing two-spinors with upper component for spin-up
electrons and lower component for missing spin-up electrons, the
$x$-depending parts are 
\begin{eqnarray} \psi_{\mathrm{inc.}}&=&  \left( \begin{array}{c} 1 \\
      0 \end{array} \right) \, e^{i k_x x} \, , \nonumber \\
 \psi_{\mathrm{a}} =\left( \begin{array}{c} 0 \\ 1 \end{array} \right)
 \, e^{i k_x x} \, , \quad
  \psi_{\mathrm{b}} &=&\left( \begin{array}{c} 1 \\ 0 \end{array}
  \right) \, e^{-i k_x x} \, ,   
 \nonumber  \\ 
\psi_{\mathrm{c} } &=& \left( \begin{array}{c} u( \theta ) \\ \eta^*
    (\theta ) v (\theta ) \end{array} \right) \, e^{i k_x x} \, ,
\nonumber  \\   
\psi_{\mathrm{d} } &=& \left( \begin{array}{c} \eta (\pi-\theta ) v
    (\pi -\theta ) \\ u(\pi - \theta ) \end{array} \right) \, e^{-i
  k_x x} \, . \label{wf} \end{eqnarray} 
where 
 \[u ( \theta ) = \sqrt{ \frac{1}{2} \left( 1 + \frac{(E^2 - |\Delta (
       \theta ) |^2 )^{1/2} }{E} \right) } \] and \[ v ( \theta ) =
 \sqrt{ \frac{1}{2} \left( 1 - \frac{(E^2 - |\Delta ( \theta ) |^2
       )^{1/2} }{E} \right) } \, ,  \]  
and $\eta (\theta ) = \Delta (\theta ) / |\Delta ( \theta)|$ is the
gap phase. For the non-unitary state (\ref{sz}) and for the spin-up
channel of the unitary state (\ref{bw2}) we have $\eta (\theta ) =
\exp ( -i \theta)$ while for the spin-down channel of the unitary
state $\eta (\theta ) =   \exp ( i \theta)$. For time-reversal
conserving singlet states $\eta (\theta )$ can be taken real and, in
particular, for $s$-wave pairing with $\eta (\theta ) =1$ we recover
the BTK results.  

  Using the boundary conditions we obtain the
  following expressions for $a$ and $b$: 
\begin{eqnarray} a_e (E, \theta ) &=& \frac{\eta^*( \theta ) v (\theta
    ) u (\pi - \theta )}{ u (\theta ) u ( \pi - \theta ) + Z^2 (\theta
    ) \, \delta(E , \theta) } \, \label{ra2}\, , \\ 
 b_e (E, \theta ) &=& \frac{(Z^2 + iZ) \, \delta (E,\theta ) }{ u
   (\theta ) u ( \pi - \theta ) + Z^2 \, \delta (E,\theta )} \,
 \nonumber  \end{eqnarray} 
with \[ \delta (E,\theta ) =   u (\theta ) u ( \pi - \theta ) - \eta^*
(\theta ) \eta (\pi - \theta) v (\theta ) v ( \pi - \theta ) \, . \] 
 $Z ( \theta )=H/(v_F \cos (\theta) )$ measures the barrier
 strength. The subscript $e$ denotes the coefficient for the incident
 electrons. 
Incoming holes (subscript $h$) with the same excitation energy $E $
and group velocity 
in the same direction $\theta$ have momenta 
opposite to those of incoming electrons. 
Performing the appropriate substitutions in (\ref{wf})
we obtain 
\begin{equation} |a_h (E,\theta ) |^2 = | a_e (E, - \theta ) |^2 \,
  . \label{eh} \end{equation} 
For the states (\ref{sz}) and(\ref{bw2}) and
the given geometry the gap function seen by the holes is the
complex conjugate of that seen by the electrons under the same angle, 
i.e. Cooper pairs with reversed angular momentum along the $z$ axis. 

\section{Results for the non-unitary state}
In Fig.\ref{ab} we show results of Andreev and normal reflection
probabilities for various angles of incidence and two barrier
strengths, $Z=0.35$ and $Z=1.5$. These curves correspond to the gapped
channel (in our convention the incoming spin-up quasiparticles) of the
non-unitary state (\ref{sz}). For incoming spin-down particles, the
superconducting gap is zero, therefore we obtain N-N junction results,
which are not shown here. 

For the unitary state (\ref{bw2}) both spin
orientations have a gapped spectrum and there is a simple relation
between spin-up and -down component.
For the spin-down electrons the behavior for given angle of incidence
$\theta$ are those of the 
spin-up electrons with $-\theta$, because of the reversed angular momentum 
of the spin-down Cooper pairs ($\eta( \theta ) \to \eta^* (\theta )$
or $k_y \to - k_y$).  
The angle-independent gap magnitude sets the energy scale for the
reflection amplitudes, but their behavior is also determined by the 
phase factors $\eta (\theta)$ occurring in the denominator of
(\ref{ra2}). The latter give rise to interesting properties. 

\subsection{Asymmetry with respect to $\theta \to -\theta$} 
\begin{figure} 
\begin{center}
\includegraphics[width=0.5\textwidth]{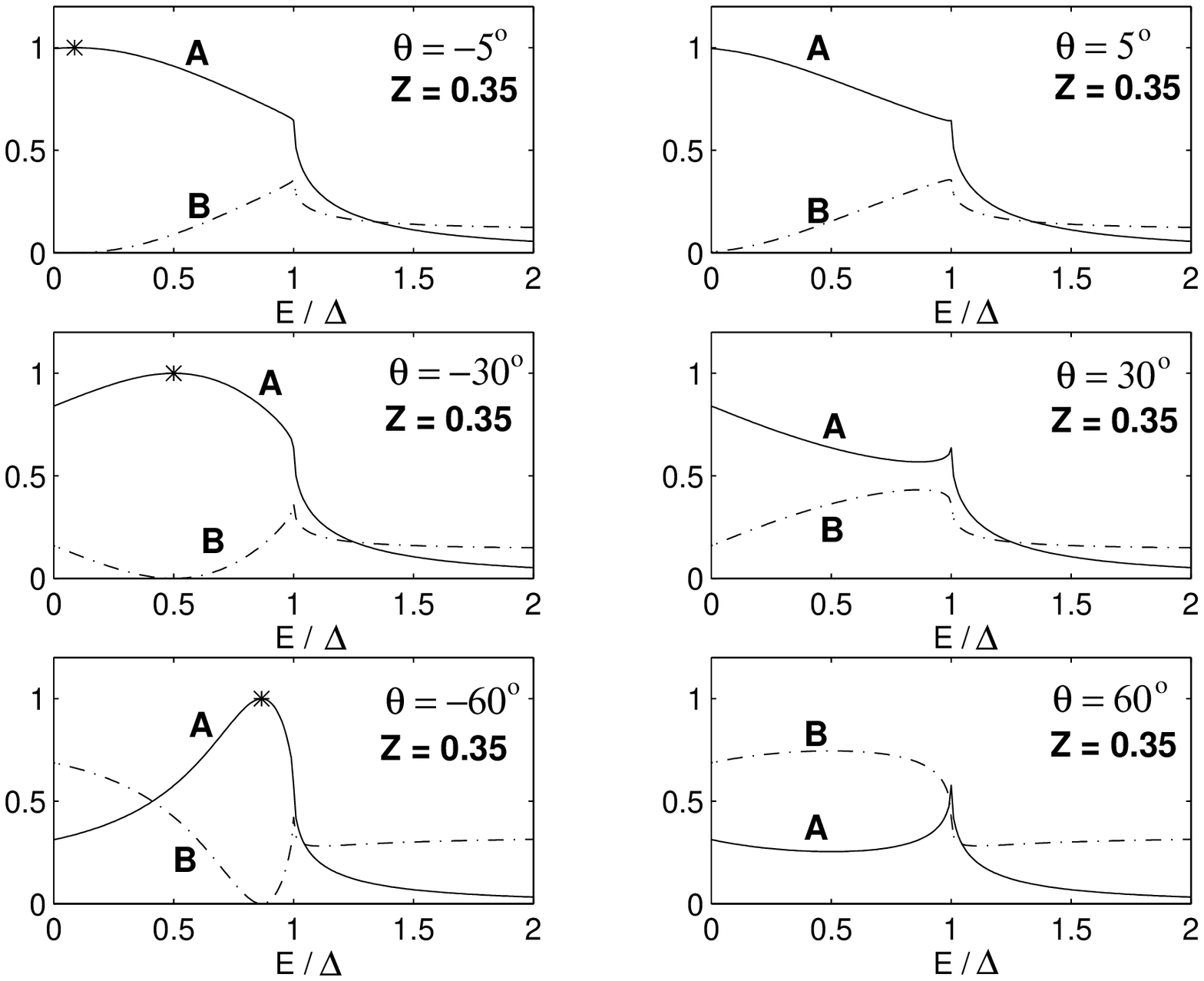}
\end{center}
\begin{center}
\includegraphics[width=0.5\textwidth]{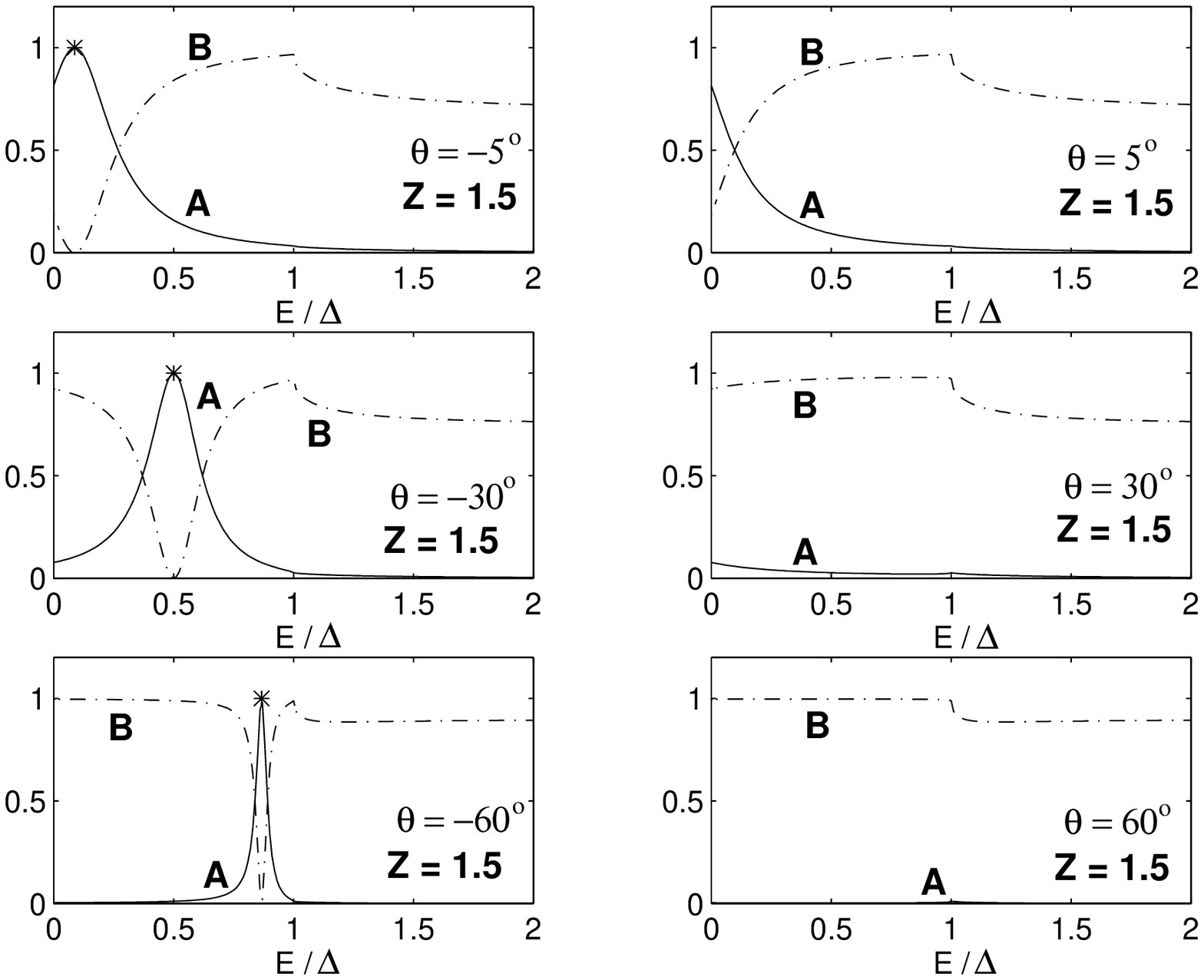}
\end{center}
\caption{Andreev ($A=|a_e(E, \theta)|^2$, solid line) and normal
  reflection ($B=|b_e(E,\theta)|^2$, dashed line) probability currents
  for an N-S boundary with non-unitary superconducting state
  (\ref{sz}) and incoming spin-up electron with energy $E$. The plots
  show the results for various angles of incidence $\theta$ and two
  different barrier strengths $Z$. The $*$ denotes the resonance energy
  $E_A = - \Delta_0 \sin \theta$. } \label{ab} \end{figure}

If we
consider the triplet states (\ref{sz}), (\ref{bw2}) and incoming particles of one spin projection (say spin-up) 
\[ \eta^* (\theta ) \eta (\pi - \theta) \not= \eta^* (- \theta ) \eta
(\pi + \theta)  \, . \] As a consequence the Andreev and normal reflection amplitudes (\ref{ra2})
are {\it asymmetric} with respect to the boundary normal. This can be seen
in Fig.\ref{ab}. Similar asymmetries have been discussed for $s+id$
superconducting gaps by Matsumoto and Shiba\cite{mash}. But note that
for the triplet states (\ref{sz}) and (\ref{bw2})  this asymmetry
occurs for arbitrary boundary orientation, because changing the
boundary orientation is equivalent to multiplying the gap function
with a constant phase. In the unitary BW state (\ref{bw2}) the
asymmetry cancels if we sum over the spin projections because  as mentioned above for
incident spin-down quasiparticles the $\eta$'s change into their
complex conjugates, which is equivalent to $\theta \to - \theta$. 
 \begin{figure}
\begin{center}
\includegraphics[width=0.5\textwidth]{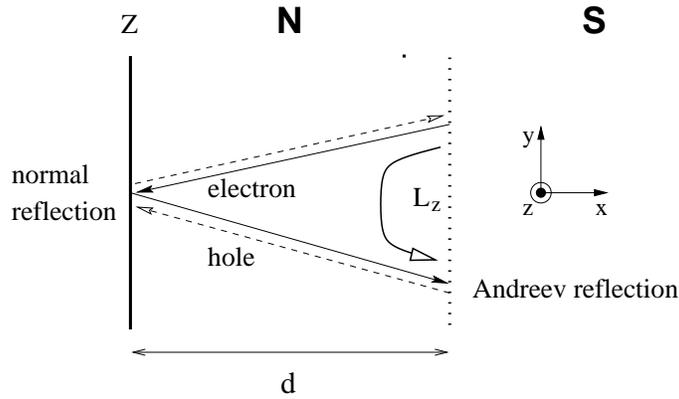} \end{center}
\caption{Formation of a bound state by constructive interference
  between normal and Andreev reflected quasiparticles. This state has
  non-vanishing angular momentum with respect to the $z$-axis.}
\label{bsfig} \end{figure}  
\subsection{Sub-gap resonances} For certain sub-gap energies and 
angles of incidence, $\delta( E, \theta)$ in the denominator of (\ref{ra2}) vanishes. This occurs under the condition
\begin{equation} \eta^* (\theta ) \eta (\pi - \theta ) =
  \frac{u(\theta) u  (\pi - \theta)}{v (\theta ) v (\pi - \theta )} \,
  . \label{bsc} \end{equation} 
Then the effect of the
boundary barrier is turned off and we obtain
$|a_e(E, \theta)|^2=1$ which corresponds to a resonance. 
The corresponding peaks in the Andreev
reflection probability become sharper with increasing $Z$, their width decreases $\propto Z^{-2}$.  At the
resonance condition  (\ref{bsc}), the transmitted wave in the superconductor is a superposition of a surface bound state (note that this expression is
only sensible for $Z \to \infty$) formed by repeated Andreev and
normal reflection (see Fig.\ref{bsfig}, consider the limit $d \to 0$)
and a $c$-type $\gamma_{\bf k}^{\dagger}$ quasiparticle, which carries
the charge $2e$ into the superconductor. Both parts of the transmitted
wave decay on a length scale $v_F \cos \theta /\sqrt{ \Delta_0^2 - E^2}$. The
condition $\delta (E, \theta) = 0 $ is equivalent to a vanishing bound
state wave function at the barrier at $x=0$. In the quasi-classical
picture  of Fig.\ref{bsfig} the bound state is formed if the phases
accumulated by the electron and hole components of the two-spinors
during one round trip through the normal layer sum up to an integer
multiple of $2 \pi$. Since  the phase shifts for the Andreev
reflections are angle-dependent, the resonance energies also vary with
$\theta$. 

In the non-unitary state (\ref{sz}) $\delta( E, \theta)$ vanishes for
$E_A= - \Delta_0 \sin \theta$, i.e. negative angles of incidence. This leads to resonances in the Andreev
amplitude, $|a_e ( E_A, \theta )|^2 =1$, at negative angles of
incidence. 
Note that the bound state has a non-vanishing angular momentum along the
$z$ axis, therefore its asymmetric position at negative $\theta$ can be
understood as a consequence of the  finite relative angular momentum of
the Cooper pairs in non-unitary states. This is in agreement with the
fact that according to (\ref{eh}) for injected holes the resonance peaks in the Andreev
amplitude change side, i.e. they occur at positive angles of incidence.
Since its angular momentum $L_z$  is fixed by the gap function, the 
bound state is only accessible for electrons with angle of incidence
$\theta$ and holes with angle of incidence $-\theta$. 
In the spin down channel of the unitary state (\ref{bw2}) the
analogous bound state with opposite $L_z$ does exist, therefore the
resonances occur 
symmetrically if we assume unpolarized incoming particles. 

The bound state condition $\delta (E, \theta ) = 0$ is essentially
equivalent to the equations recently given by Hu \cite{hu} and Tanaka
et al. \cite{tanaka} for $d$-wave and other singlet
superconductors. They report zero energy states for certain orientations  of the boundary with respect to the pair wave function. This is in contrast to the triplet states (\ref{sz}) and (\ref{bw2}) where 
the sub-gap resonances are not sensitive to the boundary
orientation in the plane. 

\subsection{Zero-bias conductance peaks}
The charge current through the NS boundary by an incoming electron is increased by an Andreev hole with probability $|a_e(E, \theta)|^2 $ and diminished by normal reflection with probability  $|b_e(E, \theta)|^2 $. Therefore at $T=0$K for the non-unitary state (\ref{sz}) and applied voltage $V$ such that we have incoming electrons from the N side, the conductance of the gapped spin-up channel is given by\cite{btk} 
\[ G^{\uparrow}_{\mathrm{NS}}(V) = e^2 \int dk_{\bot} \, v_g (k_{\bot}) \rho^{\uparrow}(eV,k_{\bot}) \, P (k_{\bot } ) \,  \left( 1  + |a_e(eV, \theta)|^2- |b_e(eV, \theta)|^2 \right)  \, . \]
Here, the reflection amplitudes have to be evaluated at energy $E=eV$ and at the angle of incidence in the plane $\theta $,  $\rho^{\uparrow}(eV, k_{\bot })$  and $v_g(k_{\bot })$ denote the spin-up density of states and  group velocity on the N side at transverse momentum $k_{\bot}$ parallel to the boundary, over which we average with probability distribution $P (k_{\bot }) $  . The latter depends on the type of experiment one is interested in. For the non-unitary state (\ref{sz}) the spin-down conductance is equal to the normal state conductance.

For $s$-wave superconducting
states (all $\eta$'s $=1$), the Andreev amplitude is largest at $E=
\Delta$. Therefore the conductance $G_{NS}$ 
 is also peaked
at gap energy and resembles the bulk density of states for low
transparency boundaries. This so called  gap-like feature is entirely
absent for the triplet states (\ref{sz}) and (\ref{bw2}).  Due to the
bound states at $E = \pm \Delta_0\sin \theta $, the Andreev amplitude
for small angles of incidence is largest around $E \approx 0$. This
leads to zero-bias conductance peaks (ZBCP)  which become increasingly pronounced when the junction 
favors smaller angles of incidence. As explained above for the
two-dimensional triplet states (\ref{sz}) and (\ref{bw2}) these ZBCPs
do not depend on the boundary orientation. Therefore they should
provide reliable evidence whether a nodeless triplet state with $\Delta (\theta ) \propto \exp (\pm i \theta )$ is
realized in Sr$_2$RuO$_4$.

Yamashiro and Tanaka \cite{yamashiro} have also reported that ZBCPs
 should occur for (100) NS boundaries with the non-unitary superconducting triplet states (\ref{sz}) and $\Delta_{\uparrow \uparrow} (\theta ) \propto \cos \theta +\sin \theta $, which has nodes at $\theta=-\pi/4$ and $\theta = 3\pi/4$ and two lobes with different signs of $\Delta( \theta )$ in between. By looking at the phase structure of the latter gap function, we expect the ZBCP
 to be strongest for a (110) boundary (one of the two lobes is directed towards the interface which implies zero-energy resonances for all angles of incidence) and a smeared-out gap-like feature and no ZBCP for ($\bar{1}$10) boundaries (lobes parallel to the interface, no sub-gap resonances). Thus such states should be experimentally distinguishable from the nodeless states (\ref{sz}) and (\ref{bw2}).

\section{Experimental distinction between unitary and non-unitary states}
\begin{figure}[h]
\begin{center} \includegraphics[width=0.5\textwidth]{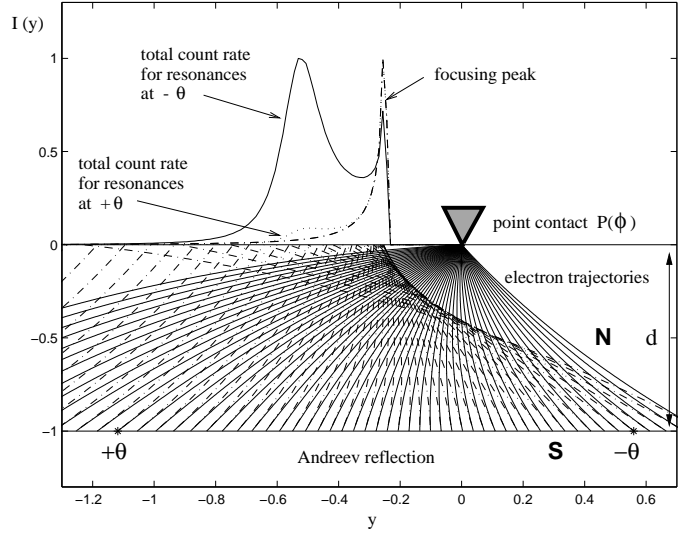} \end{center}
\caption{Scheme of a two-point-contact measurement. The three curves in
  the upper half of the plot show the distributions due to focusing
  (dashed-dotted line), and including the angular dependent Andreev
  reflection amplitude in the non-unitary state (\ref{sz}) for
  resonances at negative angle of incidence ($- \theta$, solid line)
  and resonances at positive $\theta$ (dotted line). All curves assume
  a Gaussian distribution for the injection angle $\phi$ with $\sigma
  = 20^{\circ}$. 
 }\label{pa} \end{figure}

The interplay between asymmetries and bound states can be utilized in
an experiment in order to distinguish between the
 unitary and non-unitary
states proposed for Sr$_2$RuO$_4$.  A feasible way for such
an experiment is two-point spectroscopy described by Benistant et
al.\cite{ben} .  Here, one injects unpolarized electron-like
quasiparticles through a point contact into a normal conductor of
thickness $d$ attached to the superconductor with a certain
distribution $P(\phi)$ of the injection angle $\phi$. Both the
electrons and the Andreev reflected 
holes are deflected by a magnetic field applied parallel to the $z$-axis.
The holes are now collected by a second point
contact. Obviously, experiments of this kind require a normal
conducting crystal 
of high quality where quasiparticles move ballistically.
Suppose we have many hole-collectors along the transverse
direction of the field, i.e. we are able to measure the intensity of
Andreev holes arriving at the surface of the normal conductor as a
function of the transverse coordinate $y$. 
If the probability for Andreev reflection is angle-independent,
e.g. for a conventional $s$-wave superconductor, one obtains a hole
distribution which is peaked at a minimal distance $y=y_0$ from the
injection point at $y=0$. This feature is known under the name ''electron focusing'' (EF)\cite{tsoi}.
The intensity peak at $y=y_0$ can be also
seen in Fig.\ref{pa} in the density of the 
hole trajectories arriving at the surface. It should be noted that for
the focusing to work  any voltage drop should take place at the point
contacts only, otherwise the motion of the particles would be influenced by
inhomogeneous electric fields inside the normal metal. For voltages $eV$
comparable to $\Delta_0$ the quasiparticle trajectories are
independent of the voltage drop, because in this case the velocity
deviation from $v_F$ is negligible.  

For a non-$s$-wave superconducting state, we have to take care of the
angular dependent Andreev reflection probability. This
additional factor results for the triplet states (\ref{sz}) and (\ref{bw2})
in a modification of the collected intensities due to the resonance peaks, particularly 
for higher barrier strengths $Z$, where
the Andreev reflectivity is low for energies away from the resonances.
In the presence of a magnetic field
screening currents lead in general to a change in the resonance energy\cite{fogelstroem} due to a shift in the quasiparticle spectrum by\cite{tinkham} $\Delta E_{\bf k} = 
{\bf v}_s \cdot {\bf k} = v_s k_F \sin \theta$, where ${\bf v}_s $ is the superfluid velocity. Since both transmitted quasiparticles responsible for the bound state formation have the same transverse momentum (see Fig.\ref{gapvec}) and the bound state energies have the same angular dependence $\propto \sin \theta$ as the energy shift, this only leads to a renormalization of the energy scale and the qualitative picture drawn here remains valid.  
If the weight of the quasiparticles which satisfy
the resonance condition (\ref{bsc}) is sufficiently high, they give
rise to a broader second peak  (besides the focusing peak) in the
''count rate'' at a different collector position (see Fig.\ref{pa} at
$y \approx -0.5 d$).
The magnetic field leads to an asymmetric distribution of angles of incidence
at the NS boundary. For the non-unitary state (\ref{sz}) the resonances
occur only for negative angles of incidence (e.g. at $- \theta$ in
Fig.\ref{pa}), for which the corresponding electron paths have small
injection angles $\phi$. 

\begin{figure}[h] 
\begin{center}
\includegraphics[width=0.5\textwidth]{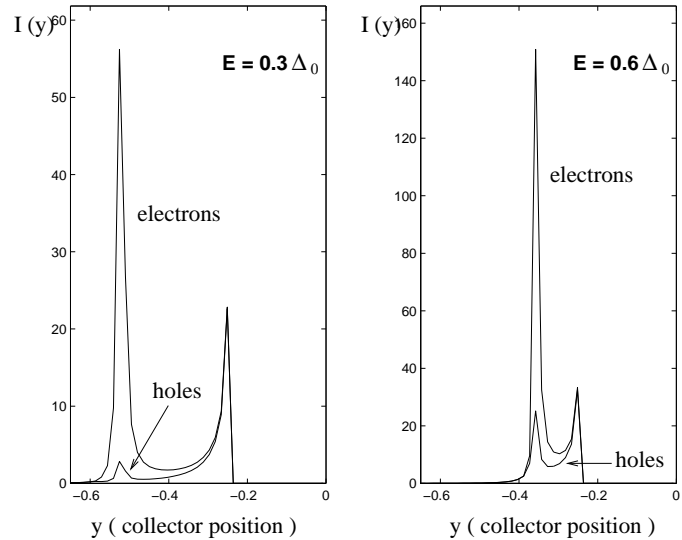}
\end{center} 
\caption{Non-unitary state: Intensities of collected Andreev particles
  versus collector 
  position for barrier strength $Z=2.5$, fixed cyclotron radius $r=4d$ and two
  energies. The 'electron' curves denote injection of electrons into the normal metal layer and are obtained from the 'hole' curves 
  by inverting the voltage between emitter and collector.}\label{cs}  
\end{figure}

 Inversion of the voltage difference between emitter and collector is
 equivalent to injecting holes instead of electrons. The injected
 holes and Andreev reflected electrons follow 
 the same paths as the electrons and reflected holes did before the
 voltage reversal. But according to (\ref{eh}) the holes face a gap
 potential with reversed angle of incidence $- \theta$, which makes
 the resonance peaks change sides for the non-unitary state (\ref{sz})
 (of course the sign of the detected voltage is also reversed). Now
 the paths corresponding to the resonance peaks at $\theta$ have large
 angle of incidence $\phi $ (see Fig.\ref{pa}).  If the angular
 probability distribution $P (\phi )$ of the injected particles is
 peaked at small injection angles this will lead to observable changes
 in the ''count rate'' of Andreev particles. In Fig.\ref{cs} we show
 the count rates versus collector position for a non-unitary
 superconducting state (\ref{sz}) 
 and two different energies of the injected
 particles. Note that the assignment of the two curves to the electron and
 hole injection depends on which one of the degenerate
 non-unitary states is realized in the superconductor. In the
 calculations, we assumed that the angular distribution $P(\phi)$  is
 Gaussian with width $\sigma = 20^{\circ}$. For increasing $\sigma$
 the asymmetry between the electron and hole injection becomes less
 pronounced. 

\begin{figure}[h]
\begin{center}
\includegraphics[width=0.5\textwidth]{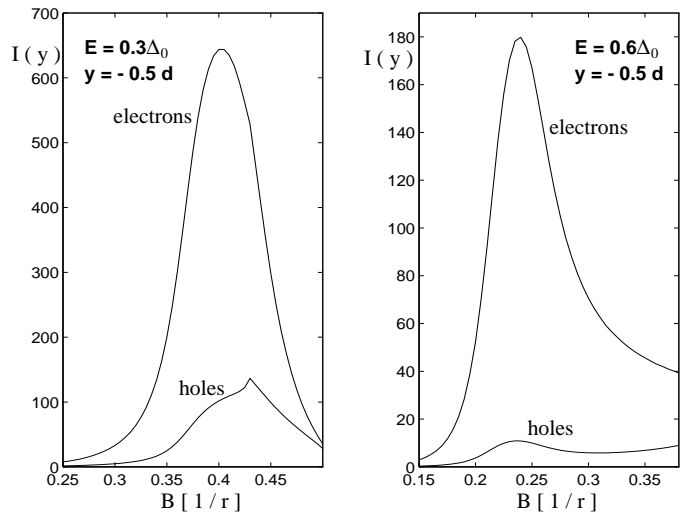} 
\end{center}
\caption{Non-unitary state: Intensities of collected Andreev particles
  versus applied 
  magnetic field for $E= 0.3 \Delta_0$ (left plot) and $E= 0.6
  \Delta_0$ (right plot), $Z=2.5$. Fixing the sign of the voltage drop
  at the emitter (i.e. electron or hole injection) corresponds to
  selecting one of the two solid lines.  The magnetic field is
  measured in $1/r$, where $r$ denotes cyclotron radius measured in
  units of the thickness of the normal conducting layer.}\label{bpasz}
 \end{figure}
\begin{figure}[h] 
\begin{center}
\includegraphics[width=0.5\textwidth]{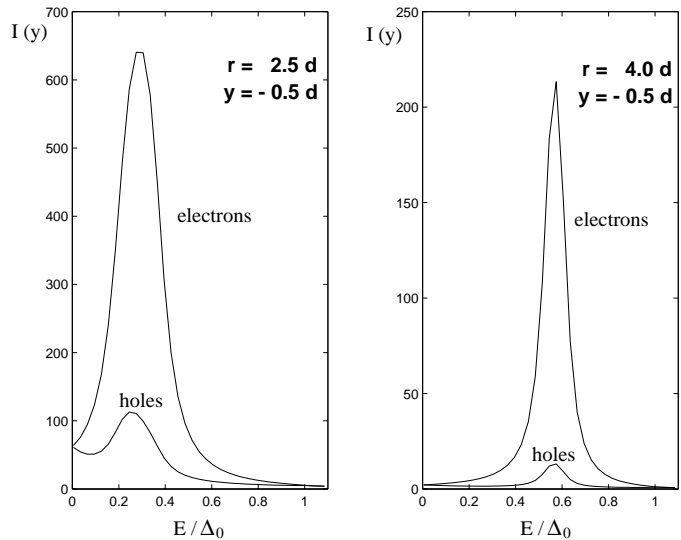}
\end{center} 
\caption{Non-unitary state: Intensities of collected Andreev particles
  versus energy $E$ of the injected particles (applied voltage) for
  fixed collector position $y$ and two magnetic field strengths,
  $Z=2.5$.}\label{epc}   
\end{figure}

 
In a real experiment one will vary the applied field instead of the
collector position. In Fig.\ref{bpasz} we plot the count rates for a
fixed collector position at energies $E=0.3 \Delta_0$ and $E=0.6
\Delta_0$, $Z=2.5$, $\sigma = 20^{\circ}$.  For collector positions
away from the the focusing peak 
the asymmetry between electron and hole injection is very pronounced
because the probability for Andreev reflection is small outside the
resonance peaks. For Sr$_2$RuO$_4$ the lower critical field along the $c$ direction is $H_{c1}(0)=11$mT\cite{hc1}. In a focusing field of $H=9$mT and using the numbers given by Benistant et. al. \cite{ben} for a silver normal conducting layer, the radius of the electron paths is $r \approx 0.88$mm. 
For example the right plot in Fig.\ref{epc}, i.e. $d=r/4 \approx 0.22$mm, corresponds to a distance of $y=0.5 d  \approx 0.11$mm between collector and injection point. All these values are in the same range as those of the experiments performed by Benistant et. al. for conventional superconductors.  
 
For the two-dimensional BW state (\ref{bw2}) we do not expect any
differences in the count rates if we reverse the voltage between the
two contacts, because for unpolarized injected particles there is no
asymmetry  with respect to $\theta \leftrightarrow - \theta$ and
consequently no difference between electron and hole injection. For
fixed voltage, we have to add the two curves denoted by ''electrons'' and ''holes'' in the plots above in
order to obtain the spin-averaged count rate for one type of quasiparticles
injected. 
\section{Summary} 
Our analysis of the properties of two typical spin-triplet pairing
states at NS interfaces has shown the existence of angular dependent sub-gap
resonances in the Andreev reflection. These resonances correspond to
bound states and are responsible for
zero-bias conductance peaks independent of the boundary orientation in
the $xy$-plane within our model. The angular dependence of the
resonance has important implications for the non-unitary state. The
reflectivity of electrons and holes (averaged over both spin
components) is not equal for the same angle of incidence. This property is
connected with the broken time reversal symmetry of the non-unitary
state. The electron and the hole probe the time reversed
pairing states. This effect could be used to identify the non-unitary
state experimentally. The EF technique explained above allows to
probe a certain range of incident angles and would reveal the broken
time reversal symmetry as an asymmetry between positive and negative
bias voltages. 

Recent investigations on high-temperature superconductors suggest that
a state with locally broken time reversal symmetry could appear for certain
orientations of the interface. This state, the so-called $ d + is
$-wave state, would lead to a splitting of the zero-bias anomaly seen
in the pure $d$-wave state for such an interface\cite{fogelstroem}. This more simple probe
does not apply in our case where for both  triplet states (3) and (5) there is
no significant qualitative difference in the I-V characteristics.

An additional interesting aspect of the non-unitary states is that one
spin orientation generates an excess current in the NS-interface while
the other one does not. Therefore, the quasiparticle current of such an
NS-interface transports not only charge but also spin. 
In this context, however, the spin-orbit coupling of
Sr$_2 $RuO$_4$ plays an
important role in determining the spin polarization direction.
Hence, a more careful consideration of the transfer of
the pseudospin states at the NS-interface is necessary.

\section*{Acknowledgments}
We would like to thank to T.M. Rice, D. Agterberg, Ch. Bruder, Y.Maeno, A.Mackenzie and A. Fauchere for many helpful discussions. One of the authors (M.S.)
is grateful for support by the Swiss Nationalfonds
(PROFIL-Fellowship).

\end{document}